# Derivation of Bell's locality condition from the relativity of simultaneity.

Casey Blood
Professor Emeritus of Physics, Rutgers University
Sarasota, FL

## Abstract
One way to deal with the fact that many versions of reality simultaneously exist in the wave function is to suppose there are hidden variables that single out one version for perception. Bell showed theoretically, and the Aspect experiment confirmed that there could be no hidden variable theory which satisfied a locality condition. We show here that the locality condition can be derived from the well-established principle of the relativity of simultaneity. Thus virtually all hidden variable theories, not just obviously local ones, are in conflict with the results of the Aspect experiment and are therefore forbidden.

**Note:** Soon after this was published, it was brought to my attention that Nicolas Gisin had earlier presented essentially the same argument, using the simultaneity of relativity and Bell's work, in arXiv:quant-ph/1002.1390 and 1002.1392. The caveats at the end of his paper on the relativistic methodology also apply to my work.

## Introduction

There can be several simultaneously existing versions of reality in the wave function (state vector) of quantum mechanics—Schrödinger's cat can be both alive and dead at the same time, for example. One proposed way to reconcile this with our perception of a single version of reality is to suppose there are 'hidden' variables that single out one version for perception [1,2]. The variables are called 'hidden' because there is presumably no direct experimental way to observe them, and in fact no evidence of their existence has ever been found. In spite of this, it is still one of the major interpretations of quantum mechanics. We will examine hidden variables from a theoretical point of view to see if it is possible in principle to construct such an underlying theory.

The primary theoretical work in this direction was done by Bell [3], with the accompanying experimental work carried out by Aspect et al [4]. In the Aspect experiment, two photons are (nearly) simultaneously emitted back-to-back, with their polarization states correlated. Bell found that if one made the reasonable assumption that the hidden variable information on the polarization states was carried locally by each photon, then a constraint was put on the correlations of the two polarizations. This constraint violated the predictions of quantum mechanics. When the Aspect experiment was done to test the inequality, the results agreed with quantum mechanics and decisively violated the Bell constraint, clear evidence that there could be no local hidden variable theory.

Our aim here is to show the generality of the locality assumption. If one uses the relativity of simultaneity in the two-photon Aspect experiment, one finds that the assumption can be justified under conditions that are weaker than the 'direct' argument of Bell. If these weaker conditions, which consist of the implications of the relativity of simultaneity, are accepted as valid, then it is apparently not possible to construct an acceptable hidden variable theory *of any kind* that reproduces the results of quantum mechanics and the Aspect experiment.

## The Bell-Aspect Experiment

In the Bell-Aspect experiment [4], two photons are simultaneously emitted with one traveling to the right and the other to the left. (One could make the same argument using electrons instead of photons.) There is a detector on each path that determines the two allowable polarization states, here called +1 and –1, of each photon. The state of the two photons is

$$|+1\rangle_L |-1\rangle_R - |-1\rangle_L |+1\rangle_R \tag{1}$$

so they have opposite polarizations (in any coordinate system).

The detectors can be rotated around the line of flight so they measure polarization along different axes. If the left one is rotated by an angle $\theta_L$ and the right by $\theta_R$, then the single-photon probabilities are (with the subscripts indicating the measured polarization)

$$\begin{aligned} P_{L+}(\theta_L) = P_{L-}(\theta_L) = 1/2 \\ P_{R+}(\theta_R) = P_{R-}(\theta_R) = 1/2 \end{aligned}. \tag{2}$$

Quantum mechanics predicts, and the Aspect experiment verified, that the two-photon correlation probabilities are:

$$\begin{aligned} P_{++}(\theta_L, \theta_R) = P_{--}(\theta_L, \theta_R) = \sin^2(\theta_L - \theta_R)/2 \\ P_{+-}(\theta_L, \theta_R) = P_{-+}(\theta_L, \theta_R) = \cos^2(\theta_L - \theta_R)/2 \end{aligned} \tag{3}$$

where the subscripts denote the readings on the left and right detectors, resp.

## Hidden Variables and Probability

There can be many versions of reality in the wave function. Quantum mechanics correctly predicts the characteristics of each version—wavelength, energy, locality and so on—but it does not tell us which version we will perceive. In the hidden variable interpretation of quantum mechanics, it is presumed there is some set of variables that determines the perceived outcome. These variables could be as simple as the position and velocity of a (conjectured) particle that follows one branch of the wave function, as in the Bohm model [1]. But we do not limit the hidden variables to the Bohm picture; they might be any sort of variables.

The hidden variables of the two photons together will be labeled $\lambda_{ph}$. It is also conceivable that the perceived outcome of a given run of the experiment will depend on the hidden variables $\lambda_{DL}, \lambda_{DR}$ of the left and right detectors. We collectively call the hidden variables $\lambda$. Each $\lambda$ determines the outcome of the measurements at each of the two detectors for every setting of the angles. There will be a density function, $\rho(\lambda)$, such that $\rho(\lambda)d\lambda$ gives the probability of the system having hidden variables in the 'volume' element $d\lambda$.

Consider the total space $\Omega_T$ of the $\lambda$ s, which are assumed here to be continuous. For a given angular setting of the two detectors, it will be divided into four non-overlapping regions,

$$\Omega_{++}(\theta_L, \theta_R),\ \Omega_{--}(\theta_L, \theta_R),$$
$$\Omega_{+-}(\theta_L, \theta_R),\ \Omega_{-+}(\theta_L, \theta_R), \qquad (4)$$

where the subscripts give the readings of the left and right detectors resp. for that particular $\lambda$. Each $\lambda$ is in one and only one of these regions. The probabilities are given as integrals over the volume elements. For example,

$$P_{++}(\theta_L, \theta_R) = \int_{\Omega_{++}(\theta_L, \theta_R)} \rho(\lambda)d\lambda = \sin^2(\theta_1 - \theta_2)/2$$
$$P_{L+}(\theta_L) = \int_{\Omega_{L+}(\theta_L, \theta_R)} \rho(\lambda)d\lambda = 1/2 \qquad (5)$$
$$\Omega_{L+}(\theta_L, \theta_R) = \Omega_{++}(\theta_L, \theta_R) + \Omega_{+-}(\theta_L, \theta_R)$$

with similar equations for the other probabilities and single-photon volume elements.

## The Locality Condition

Suppose we specify the hidden variables while the photons are in flight. Then, since the two detectors and the photons are separate physical systems, we expect the distributions of hidden variables to be separate;

$$\rho(\lambda) = \rho_{ph}(\lambda_{ph})\rho_{DL}(\lambda_{DL},\theta_L)\rho_{DR}(\lambda_{DR},\theta_R). \tag{6}$$

If we consider just the right photon, there will be regions in $\lambda$ space where that photon will be measured (by the right detector) to have polarization +1 and regions where it is measured to have polarization –1. There will be a boundary between the two regions, which we initially designate by $S_R(\lambda_{ph};\lambda_{DR},\theta_R;\lambda_{LR},\theta_L)$. The *locality condition* is that the variables connected with the detector on the left, $\lambda_{LR},\theta_L$, can not influence the measured state of the right-hand photon, and similarly for $S_L$. So we have

$$\begin{aligned} S_R &= S_R(\lambda_{ph};\lambda_{DR},\theta_R) \\ S_L &= S_L(\lambda_{ph};\lambda_{DL},\theta_L) \end{aligned}. \tag{7}$$

The surface $S_L$ divides $\Omega_T$ into two (sets of) regions, one where a measurement at the left detector gives +1 and one where a measurement gives –1. Similarly for $S_R$. Thus the two surfaces together divide $\Omega_T$ into the four regions of Eq. (4). If $\theta_L = \theta_R = \theta$ then we see from the probability rules (Eqs. (3) and (5)) that $\Omega_{++} = \Omega_{--} = \emptyset$. In that case $\Omega_T$ is divided into just two regions and so the two surfaces must coincide;

$$S_R(\lambda_{ph};\lambda_{DR},\theta) = S_L(\lambda_{ph};\lambda_{DL},\theta). \tag{8}$$

But by varying $\lambda_{DL}$ and $\lambda_{DR}$ separately, we see this implies the surfaces are independent of the hidden variables of the detectors, so we have

$$\begin{aligned} S_R &= S_R(\lambda_{ph},\theta_R) \\ S_L &= S_L(\lambda_{ph},\theta_L) \end{aligned} \tag{9}$$

(with $S_R(\lambda_{ph},\theta) = S_L(\lambda_{ph},\theta)$) and from now on we drop the photon subscript. The surface $S_R(\lambda,\theta_R)$ divides $\Omega_T$ into two separate volumes, $\Omega_{R+}(\theta_R)$, where a measurement by the right detector gives +1, and $\Omega_{R-}(\theta_R)$ where a measurement gives –1 (with a similar statement for measurements on the left).

## The Bell Inequality

We use the notation and arguments of refs. [4] and [5], which are directly based on Bell's work. Define the function $A(\lambda,\theta_L)$ to be +1 in region $\Omega_{L+}(\theta_L)$

and –1 in region $\Omega_{L_-}(\theta_L)$, and $B(\lambda, \theta_R)$ to be +1 in region $\Omega_{R_+}(\theta_R)$ and –1 in region $\Omega_{R_-}(\theta_R)$. And let

$$E(\theta) = P_{++}(\theta) + P_{--}(\theta) - P_{+-}(\theta) - P_{-+}(\theta)$$
$$S = E(\theta_a - \theta_b) - E(\theta_a - \theta_{b'}) + E(\theta_{a'} - \theta_b) + E(\theta_{a'} - \theta_{b'}) \qquad (10)$$

where the Ps are essentially defined in terms of the As and Bs in Eq. (5). Then, using the definitions of the As and Bs and algebraic manipulations based on $|A| = |B| = 1$, one can prove that

$$|S| \leq 2. \qquad (11)$$

However, if $\theta_a - \theta_b = \theta_b - \theta_{a'} = \theta_{a'} - \theta_{b'} = \pi/8$, $\theta_a - \theta_{b'} = 3\pi/8$, then quantum mechanics—Eq. (3)—implies $|S| = 2\sqrt{2}$, so there is a clear conflict between the quantum mechanical values and what is allowed by a local hidden variable theory. The Aspect experiment showed the quantum mechanical result was correct, with the hidden variable constraint of Eq. (11) being violated by more than 40 standard deviations.

## Derivation of the Locality Condition.

The locality condition is that the surface $S_R$ ($S_L$) does not depend on either the settings or the hidden variables of the left (right) detector. Instead of simply assuming, based on classical local physics intuition, that locality holds, we give here a derivation of this condition based on the relativity of simultaneity.

**Relativity.** We assume one photon in the Aspect experiment moves to the right along the +x axis and the other moves to the left along the –x axis. Both detectors are at a distance *d* from the source. There are three 'events' in this experiment; event 0 when the two photons are emitted, event 1 when the right photon wave function is detected at detector DR, and event 2 when the left photon wave function is detected at DL. We will use three reference frames; reference frame 0 is attached to the source of the two photons, reference frame A moves with velocity *v* in the +x direction and reference frame B moves with velocity *v* in the –x direction. The coordinates of event 0 are $t = 0$, $x = 0$ in all three frames. In frame 0, the coordinates of event 1 are $t = t_0 = d/c$, $x = d$, and the coordinates of event 2 are $t = t_0 = d/c$, $x = -d$.

Events 1 and 2 are simultaneous in frame 0 but not in frames A and B. To obtain the times in those frames, we use Lorentz transformations, with results

$$t_A(1) = \gamma(t_0 - vd/c^2)$$
$$t_A(2) = \gamma(t_0 + vd/c^2)$$
$$t_B(1) = \gamma(t_0 + vd/c^2) \quad . \tag{12}$$
$$t_B(2) = \gamma(t_0 - vd/c^2)$$
$$\gamma = \sqrt{1 - v^2/c^2}$$

So we see that events 1 and 2 occur at different times in the two frames. There is a time gap $\delta t = 2\gamma vd/c^2$ between events 1 and 2 in frames A and B which can be made 'large' by supposing that $v$ is nearly equal to $c$ and that $d$ is large. These Lorentz transformations in the direction of flight do not affect the angles $\theta_L, \theta_R$ or the state of polarization.

**No dependence on the opposite detector.** We now perform the experiment and consider the events in frame A. There, event 1 occurs before event 2, so when event 1 occurs, it can make no difference (in frame A) whether the detector DL is there or not. Thus the outcome, the perceived polarization in frame A, cannot depend on the $\lambda_{DL}, \theta_L$ of detector DL. But this implies there must be a surface

$$S_R = S_R(\lambda_{ph}, \lambda_{DR}, \theta_R) \tag{13}$$

that divides the regions where the $\lambda$ s are associated with a measured polarity of +1 at DR from the regions where the $\lambda$ s are associated with –1. The surface is independent of the relativistic frame because, for a given $\lambda$, the velocity of the observer should not affect the measured polarization. (See also the Appendix.) And the velocity of the reference frame certainly does not affect the $\lambda$ s—which might be, for example, the position and velocity of a reputed particle relative to fixed point 0—although their description or labeling may be different in the moving frame. Similarly in frame B we conclude there is a surface

$$S_L = S_L(\lambda_{ph}, \lambda_{DL}, \theta_L) \tag{14}$$

that divides the regions where the $\lambda$ s are associated with a measured polarity of +1 at DL from the regions where the $\lambda$ s are associated with –1. But Eqs. (13) and (14) are just the Bell locality condition of Eq. (7)!

## Conclusion and Comments.

We have shown that if relativity of simultaneity holds, then Bell's locality condition must hold in any hidden variable theory so we don't need to *assume* locality. And if Bell's locality condition holds, then Bell's argument shows that the hidden variable theory must be in conflict with the predictions of quantum mechanics and the results of the Bell-Aspect experiment. So the relativity of simultaneity plus Bell's argument and the Aspect experiment prohibit the existence of any hidden variable theory underlying quantum mechanics.

**The relativity of simultaneity.** Is there any possibility that the relativity of simultaneity is incorrect? The Lorentz transformations of Eq. (12) are based on the constancy of the speed of light and this would seem to be a secure result. In addition, elementary particle theories are all found to be relativistic. Further, relativity and the linearity of quantum mechanics imply, through group representation theory, that 'matter' (actually the wave function) has the particle-*like* properties of mass, energy, momentum and spin. This near-miraculous tie-in between relativity and the observed 'classical' properties of matter would not hold if relativity were incorrect. Thus the relativistic transformations of Eq. (12), and hence the relativity of simultaneity, seem to be on a very firm footing.

**Particles.** Does ruling out hidden variable theories also rule out the existence of particles? If one has an underlying particle theory, the variables describing the state of the particles would be the hidden variables which presumably determine the perceived outcome of the experiment. So ruling out hidden variables also seems to rule out the existence of particles.
To support this conclusion, one can show more directly that there is no evidence for particles; all the particle-like properties of matter can be derived from the properties of the wave function alone. See sec. 6 of ref. [6].

**Bohm's model.** Bohm's model provides a hidden variable theory that, in the non-relativistic regime, does exactly what is expected of it; it singles out a particular result on just the fraction of runs predicted by the $|\psi^2(x)|$ probability law. (See also ref. [7] for attempts to make Bohm's model relativistic.) So how do we reconcile that successful hidden variable model with our result? Presumably our result implies the Bohm model cannot be extended to the relativistic regime and so it cannot give a fully adequate description of the physical universe.
A slightly more detailed reason might be as follows: The mathematics of the Bohm model specifies certain non-relativistic trajectories and a non-relativistic density of trajectories. Our results apparently imply that relativistic 'particle' trajectories and/or a relativistic density cannot be constructed in a way that gives agreement with the probability law.

**Objective reality?** It is also apparently possible to use the relativity of simultaneity to rule out mathematically based collapse theories [8]. If both that result and the one given here hold up, one has ruled out any mathematically based theory in which there exists an 'objective' (that is, single-version) physical reality, whether it consists of particles, hidden variables or the single version of the collapsed wave function. That makes the interpretation of quantum mechanics interesting indeed!

## Appendix. Dependence of the surfaces on the velocity.

Suppose we assume the velocity *does* affect the surface dividing +1 from –1. Then Eqs. (13) and (14) become

$$\begin{aligned} S_R &= S_R(\lambda_{ph}, \lambda_{DR}, \theta_R, v_A), \\ S_L &= S_L(\lambda_{ph}, \lambda_{DL}, \theta_L, v_B) \end{aligned} \quad (15)$$

As in Eq. (8), we can then use the fact that the two surfaces coincide when $\theta_L = \theta_R$ to show that the surfaces are independent of the velocity.